\documentclass[aps,twocolumn,preprintnumbers,amsmath,amssymb,superscriptaddress,showpacs,longbibliography]{revtex4-2}

\usepackage[colorlinks,bookmarks=true,citecolor=blue,linkcolor=red,urlcolor=blue,citecolor=blue]{hyperref}

\usepackage{graphicx}
\usepackage{epstopdf}
\usepackage{amsmath}
\usepackage{multirow}
\usepackage{xcolor}
\usepackage{ulem}
\usepackage[version=4]{mhchem}
\usepackage{enumitem}
\usepackage{booktabs}  

\newcommand{\RNum}[1]{\uppercase\expandafter{\romannumeral #1\relax}}

\begin{document}

\title {Superconducting phase interference effect in momentum space
}

\author{Bo Zhan}
\altaffiliation{These authors contributed equally to this work.}
\affiliation{Institute of Physics, Chinese Academy of Sciences, P.O.~Box 603, Beijing 100190, China}
\affiliation{School of Physical Sciences, University of Chinese Academy of Sciences, Beijing 100049, China}
\affiliation{ByteDance Research, Fangheng Fashion Center, Beijing 100098, China}

\author{Qiang Gao}
\altaffiliation{These authors contributed equally to this work.}
\affiliation{Institute of Physics, Chinese Academy of Sciences, P.O.~Box 603, Beijing 100190, China}

\author{Runze Chi}
\altaffiliation{These authors contributed equally to this work.}
\affiliation{Institute of Physics, Chinese Academy of Sciences, P.O.~Box 603, Beijing 100190, China}
\affiliation{School of Physical Sciences, University of Chinese Academy of Sciences, Beijing 100049, China}

\author{Yiwen Chen}
\affiliation{Institute of Physics, Chinese Academy of Sciences, P.O.~Box 603, Beijing 100190, China}
\affiliation{School of Physical Sciences, University of Chinese Academy of Sciences, Beijing 100049, China}

\author{Lin Zhao}
\affiliation{Institute of Physics, Chinese Academy of Sciences, P.O.~Box 603, Beijing 100190, China}

\author{Dingshun Lv}
\affiliation{ByteDance Research, Fangheng Fashion Center, Beijing 100098, China}

\author{Xingjiang Zhou}\email{XJZhou@iphy.ac.cn}
\affiliation{Institute of Physics, Chinese Academy of Sciences, P.O.~Box 603, Beijing 100190, China}
\affiliation{School of Physical Sciences, University of Chinese Academy of Sciences, Beijing 100049, China}
\affiliation{Songshan Lake Materials Laboratory, Dongguan, 523808, China.}

\author{Tao Xiang}\email{txiang@iphy.ac.cn}
\affiliation{Institute of Physics, Chinese Academy of Sciences, P.O.~Box 603, Beijing 100190, China}
\affiliation{School of Physical Sciences, University of Chinese Academy of Sciences, Beijing 100049, China}

\date{\today}

\begin{abstract}

Detecting the phases of the superconducting order parameter is pivotal for unraveling the pairing symmetry of superconducting electrons. Conventional methods for probing these phases have focused on macroscopic interference effects, such as the Josephson effect. However, at the microscopic level, phase interference effects within momentum space have remained elusive due to the inherent difficulty of extracting phase information from individual momentum points. By introducing the hybridization effect between a primary band and its replica bands arising from density wave orders or other interactions, we uncover a novel superconducting phase interference effect at the intersection points on the Fermi surfaces of these bands. This effect elucidates the remarkable anomalies recently observed in the single-particle spectral function through angle-resolved photoemission spectroscopy (ARPES) in $\text{Bi}_2\text{Sr}_2\text{Ca}\text{Cu}_2\text{O}_{8+\delta}$ (Bi2212) superconductors. It can also emerge in twisted junctions of superconductors with coherent tunneling, offering an alternative framework for probing the relative superconducting phase through twisted superstructures. 

\end{abstract}
\maketitle


\section{Introduction}

 Phase-sensitive measurements provide crucial insights into the nature of high-T$_c$ and other unconventional superconductors, particularly regarding their pairing symmetry and exotic excitation properties \cite{2022Xiang}. The most notable phase-sensitive effect in unconventional superconductors is observed in Josephson junctions and superconducting quantum interference devices (SQUID)~\cite{PhysRevLett.71.2134,PhysRevLett.74.797}. In a Josephson junction involving an unconventional superconductor, the current-phase relationship can exhibit a $\pi$-phase shift, indicative of non-$s$-wave pairing symmetry~\cite{PhysRevB.36.235,doi:10.1143/JPSJ.61.4283}. This so-called $\pi$-junction effect arises from the sign change of the superconducting gap function along different directions, independent of the gap amplitude.

 By fabricating Josephson junctions with different orientations relative to the crystal axes of an unconventional superconductor, phase-sensitive tests can probe the gap nodes and the symmetry of the order parameter. For example, in a $d$-wave superconductor, a corner $\pi$-junction may display maximum supercurrent at half-integer flux quantum values \cite{PhysRevLett.71.2134}. Furthermore, a spontaneous half-quantum flux can emerge in a ring with an odd number of $\pi$-junctions, such as in tricrystal Josephson rings \cite{RevModPhys.72.969, PhysRevLett.73.593}. These two fundamental effects are instrumental in identifying the $d$-wave pairing symmetry of cuprate superconductors.

 Another phase-sensitive effect is the emergence of the so-called Andreev-bound state at the interface of unconventional superconductors, resulting in characteristic zero-bias conductance peaks \cite{PhysRevLett.89.177001, PhysRevLett.80.1296,PhysRevLett.74.3451,Kashiwaya_2000}. These states are sensitive to the phase of the gap order parameter and can be used to probe the pairing symmetry. Additionally, unconventional superconductors can host half-integer flux vortices~\cite{PhysRevLett.76.1336}. It is again a phase-sensitive effect detectable using SQUID. The magnetic resonance peaks or neutron spin resonance observed in neutron scattering spectroscopy can also be utilized to identify unconventional pairing symmetries~\cite{ROSSATMIGNOD199186,PhysRevLett.85.3261, PhysRevB.78.140509,Song_2016}. 

Some unconventional superconductors break time-reversal symmetry, leading to spontaneous current loops or magnetic moments~\cite{sigrist_time-reversal_1998,mielke2022time}. In topological superconductors, phase-sensitive phenomena can give rise to the Majorana bound states~\cite{Rice_1995,RevModPhys.83.1057,PhysRevLett.105.177002}, which hold significant promise for quantum computing applications. 

Quasiparticle interference (QPI) occurs when quasiparticles scatter off impurities, defects, or boundaries in a superconductor, creating interference patterns in the local density of states~\cite{PhysRevB.67.020511}. These patterns, measured using Scanning Tunneling Microscopy (STM), provide valuable information about the electronic structure and superconducting gap symmetry~\cite{PhysRevLett.85.1536}. Combining QPI with an applied magnetic field leads to a new approach to probe the superconducting pairing symmetry. By comparing QPI patterns obtained with and without a magnetic field, insights can be gained into the gap structure and the nature of the pairing symmetry~\cite{doi:10.1126/science.1166138}. This approach has been successfully applied to cuprate and iron-based superconductors~\cite{doi:10.1126/science.1166138,doi:10.1126/science.1187399}.

However, detecting the superconducting gap phase remains challenging generally \cite{RevModPhys.72.969, RevModPhys.84.1383}, as most measurement observables or response functions are sensitive only to the gap magnitude, not its phase. This difficulty is especially pronounced in probing any physical effect associated with the pairing phase in momentum space. A Cooper pair consists of two electrons with opposite momenta, and detecting its phase in momentum space requires an interaction capable of coupling to and altering this phase at a specific momentum. Meeting this requirement is challenging. This is why ARPES, although highly effective at probing the gap magnitude by measuring the single-particle spectral function, is not sensitive to the pairing phase~\cite{PhysRevLett.70.1553,RevModPhys.75.473,PhysRevB.54.R9678}.

One might naively think that in a multiband system, superconducting electrons could couple at intersection points on the Fermi surfaces of two energy bands, leading to experimentally detectable effects. The Fermi surfaces are relevant because superconducting pairing occurs at a much lower energy scale than the bandwidth. However, careful examination reveals that this would not work. If the two primary bands couple strongly, they will hybridize to form two new bands with an energy gap significantly larger than the superconducting gap. This hybridization effectively removes the crossing point, causing the system to behave like a one-band system at that specific momentum. Conversely, if the intersection point persists, protected by certain symmetries, the hybridization between the two bands vanishes and cannot generate any detectable interference effect.

 The discussion above suggests that to create an interaction that couples the superconducting phase at a specific momentum point, the two bands intersecting at the Fermi surface must hybridize, but not so strongly as to open a significant band gap at the intersection. Here, we point out that this seemingly complex condition can be achieved at least in the following two cases:
\begin{enumerate}[label=(\arabic*)]
    \item If the second band, which couples with the first (referred to as the main band), is induced by an interaction from the first with an energy scale significantly smaller than the Fermi energy but comparable to the superconducting pairing interaction. Such coupling occurs when the material exhibits charge density waves (CDW), spin density waves (SDW), pair density waves (PDW), lattice modulation, or other low-temperature instabilities. These instabilities commonly exist in copper-oxide, iron-based superconductors, and other correlated materials. For instance, Bi2212 exhibits an incommensurate lattice superstructure modulation along the $b^\star$ direction \cite{gao_incommensurate_1988,Withers_1988, EIBL1991419,HEINRICH1994133,PhysRevB.54.4310,PhysRevB.100.241112,Ding_1996}, leading to the formation of induced replica bands.
    
    \item If the two bands belong separately to the two superconductors in a Josephson junction and the junction is coupled predominantly by coherent tunneling that preserves momentum, a significant interaction between the two pairing gaps will be present at the intersection points, leading to a phase interference effect. This kind of coupling may occur in a twisted Josephson junction of Bi2212 superconductors with a clean interface \cite{PhysRevX.11.031011,PhysRevB.65.140513,PhysRevB.70.094517,science.abl}.
\end{enumerate}
 In both cases, we will demonstrate that the interplay between two bands can lead to a superconducting phase interference effect, revealing information about the gap signs. This interference is most pronounced at the intersection points of two weakly hybridized bands, where gapless excitations can arise from the coupling between superconducting phases and specific density waves, or through coherent tunneling in a twisted Josephson junction. Such momentum-space phase interference effect has already been observed in a recent ARPES measurement for Bi2212 at the intersection between the primary band and a replica subband induced by lattice superstructure modulation \cite{gao_arpes_2024}. This observation provides the first experimental evidence supporting the theoretical framework presented below.

\section{Phase interference effect}
\label{Sec:DW_modulation}

\subsection{Model Hamiltonian}

Below, we examine a superconducting state coupled with a CDW, SDW, PDW, or lattice superstructure modulation characterized by a wave vector $Q$ to illustrate the theory of superconducting phase interference effects in momentum space. This example highlights how the interaction between superconducting and density wave states can lead to observable phase interference effects, providing insight into the underlying gap signs. Moreover, our theoretical framework is versatile and can extend to superconducting systems influenced by other types of interactions. For instance, hybridization effects induced by structural modulations can also be analyzed within this framework. 

The Hamiltonian reads
\begin{eqnarray}
    H &=& H_0+ H_Q , \label{eq:H_Q} \\
    H_0 & = &   \sum_{k\sigma} \varepsilon_{k} c_{k\sigma}^\dagger c_{k\sigma} + \sum_k \left(\Delta_{k} c_{k\uparrow}^\dagger c_{-k\downarrow}^\dagger+\text{h.c.}\right) , \label{eq:H0} 
\end{eqnarray}
 where $c_{k\sigma}$ is the electron annihilation operator with momentum $k$ and spin $\sigma$. $H_0$ is the BCS mean-field Hamiltonian,  $\varepsilon_k$ is the energy dispersion of electrons, and $\Delta_{k}$ is the momentum-dependent gap order parameter. For an isotropic $s$-wave pairing state, $\Delta_{k}=\Delta$, independent of momentum $k$. For a $d_{x^2-y^2}$-wave pairing state, $\Delta_k \propto (\cos k_x- \cos k_y )$. $H_Q$ represents the hybridization induced by a density-wave order, a lattice modulation, or other interactions  characterized by a wave vector $Q$. Specifically, the following four cases are examined:
 \begin{enumerate}[label=(\arabic*)]
     \item CDW or lattice superstructure modulation
        \begin{equation}
        H_Q =\sum_{k\sigma}\left(V_kc_{k+Q,\sigma}^\dagger c_{k\sigma}+\text{h.c.} \right) 
        \end{equation}  
     \item SDW
        \begin{equation}
        H_Q =\sum_{k\sigma}\left(\sigma V_kc_{k+Q,\sigma}^\dagger c_{k\sigma}+\text{h.c.} \right) 
        \end{equation} 
     \item Spin-singlet PDW (PDW-\RNum{1})
        \begin{equation}
            H_Q = \sum_k V_k\left(c_{k+Q,\uparrow}^\dagger c_{-k,\downarrow}^\dagger - c_{k+Q,\downarrow}^\dagger c_{-k,\uparrow}^\dagger \right) + \text{h.c.} 
        \end{equation}
     \item Spin-triplet PDW (PDW-\RNum{2})
        \begin{equation}
            H_Q =\sum_k V_k\left(c_{k+Q,\uparrow}^\dagger c_{-k,\downarrow}^\dagger + c_{k+Q,\downarrow}^\dagger c_{-k,\uparrow}^\dagger \right) + \text{h.c.} 
        \end{equation}        
 \end{enumerate}
 where $V_k$ is the hybridization constant. In order to preserve time-reversal symmetry, a $H_{-Q}$  term is also included in the Hamiltonian for PDW cases. In a small region near the intersection points, which are the primary focus of our theory, $V_k$  can be assumed to be independent of $k$. 

 The momentum-space interference effect revealed by the Hamiltonian (\ref{eq:H_Q}) also works in twisted junctions when the interjunction coupling of electrons is weak and predominately governed by momentum-conserving coherent tunneling processes. A detailed discussion of this quantum phase interference effect in twisted junctions will be presented in Sec. \ref{Sec:twisted}. 
 
 Starting from an energy band characterized by $\varepsilon_k$, referred to as the main band, a sequence of induced sub-bands is generated by iteratively applying $H_Q$ to this band. If $Q$ is commensurate, then $Q= ( 2 \pi p_x, 2 \pi p_y)$ with $p_x$ and $p_y$ rational numbers. For simplicity in the discussion given below, we assume $p_x = 1/ N_x$ and $p_y = 1/N_y$ with $N_x$ and $N_y$ integer numbers. In this case, the induced band will cyclically revert to the original one with a periodicity of $N$, which is the least common multiple of $N_x$ and $N_y$ (in general, $N_x$ and $N_y$ can be any rational numbers). Consequently, beginning at a momentum point $k$, the points generated by applying $H_Q$ multiple times, {\it i.e.} \{$k + nQ \, \vert \, n=0, 1, \cdots, N-1$\}, will form a closed loop. From lattice momentum conservation, it can be shown that applying $H_Q$ $N-1$ times to the main band is equivalent to applying $H_{-Q}$ once to the main band. The first sub-bands contain all the bands generated by applying $H_Q$ or $H_{-Q}$ once to the main band. Similarly, the $n$th sub-bands contain all the bands generated by applying $H_Q$ or $H_{-Q}$ $n$ times to the main band. At the intersection points of the Fermi surfaces between the main and induced sub-bands, the gap parameters in unconventional superconducting states may change sign, either matching or opposing each other.

\begin{figure*}[t]
\includegraphics[width=0.9\linewidth]{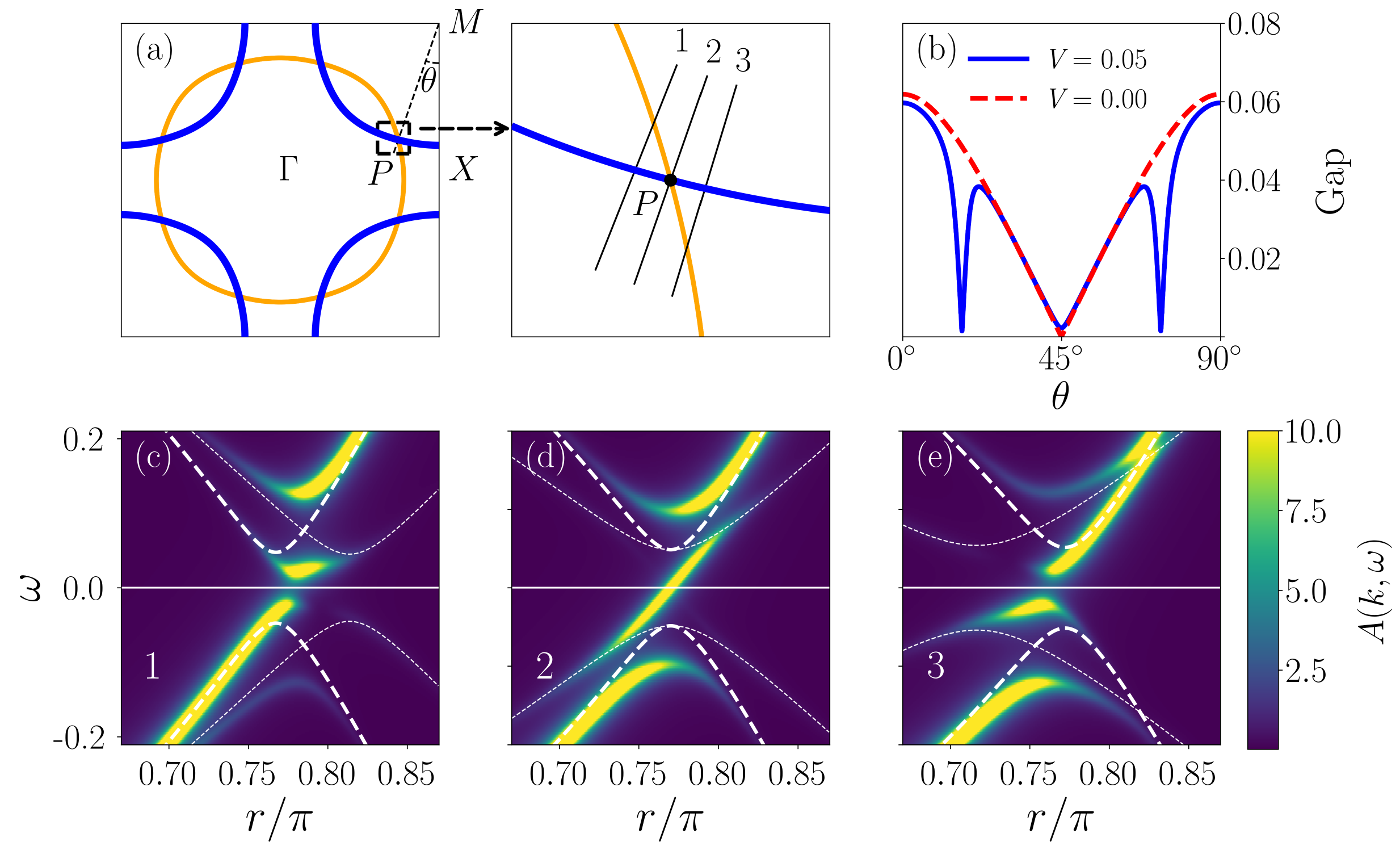}
\caption{
 (a) Fermi surfaces of the main band (blue) and the first subband (orange) with $Q = (\pi, \pi)$. The panel on the right-hand side of (a) shows an enlarged view of the Fermi surfaces around the intersection point $P$ at $(r, \theta )= (0.78\pi, 19.0^\circ) $ in the polar coordinate whose origin is at $M = (\pi, \pi) $. The three black lines represent three radial paths at angles $\theta = 21.4^\circ$, $19.0^\circ$, and $16.7^\circ$, labeled as ``1", ``2", and ``3", respectively. (b) Superconducting energy gap as a function of $\theta$ on the main-band Fermi surface (blue), compared with a pure $d$-wave gap without the main and sub-band interference (dashed red curve). (c-e) Spectral functions along the three radial paths in a $d_{x^2-y^2}$-wave pairing state. Thick (Thin) dashed white curves indicate quasiparticle energy dispersions of the main band (subband) before considering their phase interference effect along the three paths. An interference-induced node emerges at the intersection point shown in (d), while no additional gap nodes are observed along the other two paths, as illustrated in (c) and (e). Although the spectral weights at the Fermi level in (c) and (e) appear finite, they arise solely from the broadening of energy levels introduced by the parameter $\eta$ in Green's function. The parameters used are $(V, \Delta) = (0.05, 0.07)$ and $\eta = 0.01$. The white solid line represents the Fermi energy.
 }
\label{fig:Fig_Q2}
\end{figure*}

\begin{figure*}[t]
\includegraphics[width=0.9\linewidth]{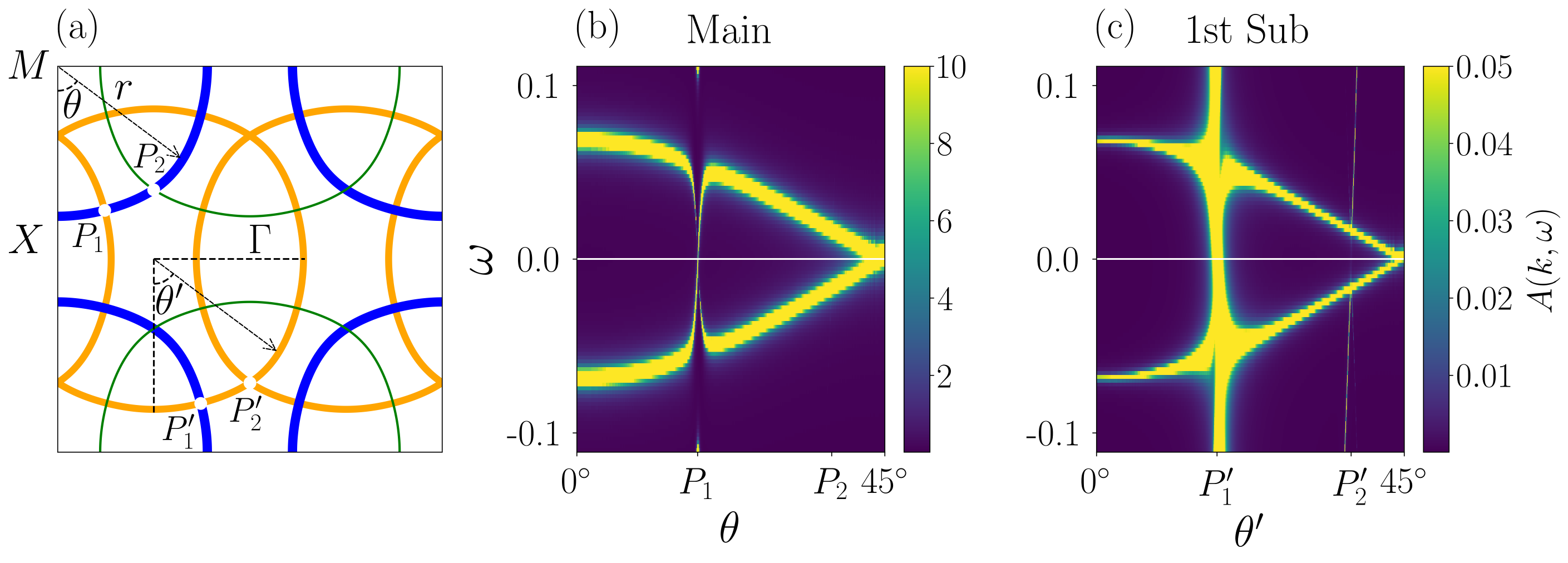}
\caption{ 
 (a) Fermi surfaces of the main band (blue), the first subband (orange), and the second subband (green) induced by $H_Q$ with $V_\sigma = V$ and $Q=( \pi /2,\pi)$. $P_1$ ($P^\prime_1$) is an intersection point between the main and first subband Fermi surfaces, where the $d_{x^2-y^2}$ pairing gaps of the main and first subband electrons show opposite signs. $P_2$ is an intersection point between the main and second subband Fermi surfaces, where the $d_{x^2-y^2}$ pairing gaps of the main and second subband electrons show the same sign. (b) Single-particle spectral function along the Fermi surface of the main band from $\theta = 0$ to $\pi /4$ on the second quadrant of the Brillouin zone. (c) Single-particle spectral function along the Fermi surface of the first sub-band from $\theta^\prime = 0$ to $\pi /4$, where $\theta^\prime$ is the polar angle defined in the framework centered at $(-\pi,-\pi) + Q$. The parameters used are $(V, \Delta)=(0.05,0.07)$ and $\eta = 0.0005$. The white solid line represents the Fermi energy.}
\label{fig:Fig1_Q4}
\end{figure*}

 In the discussion below, we assume that the energy dispersion of electrons is given by
\begin{eqnarray}
     \varepsilon_{k} &= &-2t_1\left(\cos k_x+\cos k_y \right)-4t_2\cos k_x \cos k_y \nonumber \\
     &&-2t_3\left(\cos 2k_x +\cos 2k_y \right) \nonumber \\
     &&-4t_4\left(\cos 2k_x \cos k_y +\cos k_x \cos 2k_y \right) - \mu , \label{eq:energy}
\end{eqnarray}
 where $t_i$ ($i=1, 2, 3, 4$) is the $i$th nearest neighbor hopping integral on the square lattice, and $\mu$ is the chemical potential. The parameters obtained by fitting the Fermi surface and band structures of Bi2212~\cite{gao_arpes_2024}, {\it i.e.} $(t_2,t_3,t_4,\mu) / t_1 =(-0.342, 0.11, -0.087,0.64)$, are used. $t_1$ is set as the unit of energy unless otherwise specified. However, it should be noted that the physics described below applies generally and is not dependent on the specific band structure.  
 
 As an illustration, let us first consider a $d_{x^2-y^2}$-wave pairing system with a band modulation induced by a CDW interaction and $Q=(\pi, \pi)$ where $N_x = N_y = 2$. Since the periodicity $N=2$, only the first subband is induced. The $Q=(\pi,\pi)$ modulation usually comes with the $(\sqrt{2},\sqrt{2})$ superstructure reconstruction. This type of CDW or superstructure modulation was observed in overdoped bismuth-based cuprates like Bi2223~\cite{Zou_2022} and Bi2201~\cite{Li_2018}. In this case, the Hamiltonian can be rewritten as 
\begin{equation} \label{Eq:H(k)}
    H = \frac{1}{N}\sum_{k} \psi_{k}^\dagger H(k) \psi_{k}
\end{equation}
where $\psi^\dagger_{k}=(c^\dagger_{k\uparrow},c^\dagger_{k+Q,\uparrow},c_{-k\downarrow},c_{-k-Q,\downarrow}) $ is the Nambu vector of electrons and 
\begin{equation}
    H (k) = \left(
    \begin{array}{cccc}
    \varepsilon_k & V & \Delta_k & 0 \\
    V^* & \varepsilon_{k+Q} & 0 & \Delta_{k+Q} \\
    \Delta_k^* & 0 & -\varepsilon_k & -V \\
    0 & \Delta_{k+Q}^* & -V^* & - \varepsilon_{k+Q} 
    \end{array}
    \right) ,
\end{equation}
which is particle-hole symmetric with $V$ the hybridization constant. The eigenvalues of this 4$\times$4 Hamiltonian are
\begin{equation}
    E_k = \pm\frac{1}{\sqrt{2}}\sqrt{\varepsilon_k^2+\varepsilon_{k+Q}^2+\Delta_k^2+\Delta_{k+Q}^2+2V^2\pm M_k} ,\label{Eq:Ek}
\end{equation}
where
\begin{eqnarray}
        M_k & = &\left\{ 4V^2 \left[ \left( \Delta_k - \Delta_{k+Q} \right)^2 + \left( \varepsilon_k + \varepsilon_{k+Q} \right)^2 \right] \right. \nonumber \\
        &&\left. +  \left( \Delta_k^2 - \Delta_{k+Q}^2 + \varepsilon_k^2 - \varepsilon_{k+Q}^2 \right)^2 \right\}^{\frac{1}{2}} .
\end{eqnarray}
At the intersection points of the Fermi surface, where $\varepsilon_k = \varepsilon_{k+Q} = 0$ and $\Delta_{k+Q} = -\Delta_k$ in the $d_{x^2-y^2}$-wave pairing state, the eigenvalues simplify to $E_k = \pm(\Delta_k \pm V)$. An interesting feature is that the quasiparticle excitation gaps vanish when the hybridization constant exactly equals the gap amplitude, $V = \pm \Delta_k$, introducing a new node in addition to the conventional $d_{x^2-y^2}$-wave gap nodes along the zone-diagonal directions. This additional gap node arises purely from the phase interference effect. Similar phenomena are observed for other combinations of density waves and superconducting pairing symmetries.

        


\begin{table}[bh]
\caption{Eigenenergies at the intersection points of the Fermi surfaces with $Q=(\pi,\pi)$ for different pairing symmetries and density wave states.}
\label{tab:eigenvalues}
\centering
\begin{tabular}{ccc}
\toprule[1.5pt] 
 & \multicolumn{2}{c}{Density Waves} \\
\cmidrule(lr){2-3} 
Pairing Symmetry & CDW/PDW-\RNum{2} & SDW/PDW-\RNum{1} \\
\midrule[1pt] 
D-wave & $E_k = \pm(\Delta_k\pm V)$ & $E_k = \pm\sqrt{\Delta_k^2+V^2}$ \\
S-wave & $E_k = \pm\sqrt{\Delta_k^2+V^2}$ & $E_k = \pm(\Delta_k\pm V)$ \\
\bottomrule[1.5pt] 
\end{tabular}
\end{table}

Table~\ref{tab:eigenvalues} presents the eigenenergies at an intersection point on the Fermi surface for different superconducting pairing symmetries and density waves. Two distinct gap behaviors are identified. The first behavior, $E_k=\pm (\Delta_k \pm V)$, is observed in the d-wave pairing state hybridized with a CDW. The minimal excitation gap decreases and becomes zero when $V=±\Delta_k$. The second behavior, $E_k =\pm\sqrt{\Delta_k^2+V^2} $, occurs in the $d$-wave pairing with an SDW order, where hybridization increases the energy. An intriguing observation is that the interference effects in PDW-II and PDW-I mirror those observed in the CDW and SDW cases.

 Hence, the gap structure is strongly influenced by the relative phases of the pairing order parameters at the intersection points of the Fermi surfaces between two weakly coupled bands. This momentum-resolved interference effect is important to understanding experimental results observed near these intersection points, which will be further explored in subsequent sections. Similar gap characteristics have been predicted and observed in iron-based superconductors~\cite{PhysRevB.80.100508}, where conventional s-wave pairing in the presence of an SDW leads to gap nodes at folded Fermi surface intersections, while the $s_\pm$ pairing does not. This observation aligns well with our theory since the $ s_\pm$ pairing gap exhibits opposite signs at the intersection points, analogous to the d-wave scenario discussed above.

In general, for a system with a hybridization vector $Q = (2\pi / N_x, 2\pi / N_y )$ and periodicity $N$, we can always represent the Hamiltonian in the form given by Eq. \eqref{Eq:H(k)}. However, $H(k)$ is now a $2N \times 2N$ matrix, and $\psi_k$ is a $2N$-dimensional Nambu vector that includes all independent momentum points generated by $H_Q$ starting from $k$. Since the interference effects in PDW-II and PDW-I resemble those in the CDW and SDW states, we will focus the discussion on the latter two cases below.

\subsection{Spectral function}

 By diagonalizing $H(k)$, we can determine the renormalized spectral function of superconducting quasiparticles, defined by the imaginary part of the retarded Green’s Function $G^R(k, \omega)$
\begin{eqnarray}
    A(k,\omega) &=& -\frac{1}{\pi}\text{Im}G^R_{11}(k,\omega) \label{eq:A(k,w)} ,\\
    G^R(k,\omega) &= & \frac{1}{\omega-H(k)+i\eta} ,
\end{eqnarray}
 where $\eta$ is a small broadening parameter that transforms a $\delta$-function into a Lorentzian peak.  In the superconducting state, $A(k,\omega)$ measures the superconducting quasiparticle weight. 

 As an illustration, let us first consider a $d_{x^2-y^2}$-wave pairing system with a CDW hybridization vector $Q=(\pi, \pi)$ whose Hamiltonian is defined by Eq.~\eqref{Eq:H(k)}.  Fig.~\ref{fig:Fig_Q2}a shows the Fermi surfaces of the main band and the first subband, with two intersection points in each quadrant of the Brillouin zone. For this particular system, the gap order parameters of the main band exhibit opposite signs to those of the subband on all the intersection points. 
  
  Fig.~\ref{fig:Fig_Q2}b shows the angle dependence of the superconducting energy gap along the main Fermi surface. Comparing this with the energy gap without considering the interference effect, we find significant modifications around the intersection points. It shows that interference between the main band and the first subband can substantially alter the gap structure around each intersection point when the energy gaps of these two bands have opposite signs. This interesting momentum-space phase interference effect was first reported in Ref.~\cite{gao_arpes_2024}. 
  
   Fig.~\ref{fig:Fig_Q2}c -e illustrates the excitation spectra along three chosen paths near the intersection point $P$, as displayed in the right panel of Fig.~\ref{fig:Fig_Q2}a. A comparison with the bare quasiparticle energy dispersions of the main band and the induced subband, shown by the thin white dashed curves in Fig.~\ref{fig:Fig_Q2}c -e, reveals significant changes in the spectra due to the coupling between these two bands. This coupling induces a pronounced renormalization of the spectral function around the intersection point. Notably, the superconducting energy gap completely vanishes at this intersection.

 The example above highlights the essential features of the superconducting phase interference effect. Below, we show that this effect occurs not only in $d$-wave pairing states but also in $s$-wave and other pairing states. In $d$-wave and other unconventional pairing systems, significant phase interference arises when the superconducting energy gaps of the main and induced bands have opposite signs at their intersection, coupled by a CDW or CDW-like interaction. Conversely, in $s$-wave pairing systems where the gap parameters do not change sign, pronounced phase interference occurs when an SDW or SDW-like interaction couples the main and induced bands.

 To gain a more comprehensive understanding of the phase interference effect, we now consider a more general system characterized by a coupling vector $Q$ that is not along the diagonal direction. Specifically, we examine a system with $Q=(\pi /2, \pi)$. We analyze both $d_{x^2-y^2}$- and isotropic $s$-wave pairing states, coupled with either CDW (where $V_\sigma = V$) or SDW (where $V_\sigma = V\sigma$). In this case, as $N_x=4$ and $N_y=2$, and their least common multiple $N=4$, besides the first subbands, the second subbands can also be induced by $H_Q$. Fig.~\ref{fig:Fig1_Q4}a depicts the Fermi surfaces of the main band along with those of the induced first and second subbands. The main band has now more intersection points with the subbands on the Fermi surfaces.

 Fig.~\ref{fig:Fig1_Q4}b -c depicts the spectral function of this system in the $d_{x^2-y^2}$-wave pairing state coupled with a CDW interaction. As expected, the spectrum along the Fermi surfaces of the first subband mirrors that of the main band, although the spectral weight of the first subband is significantly reduced compared to the main band. At the intersection points $P_1$ and $P_1^\prime$ between the main and first subbands, the gap signs change, resulting in significant modifications to the spectral function, similar to the $Q=(\pi, \pi)$ case. Notably, the superconducting energy gaps at these intersection points completely vanish, indicating the emergence of new gap nodes induced by phase interference. 
 
 The intersection point $P_2$ arises from the second-order hybridization effect associated with $2Q$, which is too weak to markedly modify the spectral function predominantly contributed by the main band. Additionally, different first subbands in this system, derived from the $Q$ and $-Q$ terms, can intersect. At their intersection point $P_2^\prime$, the gap parameters also change sign, and the corresponding phase interference effect is noticeably weak, though discernible upon close examination of Fig.~\ref{fig:Fig1_Q4}c.

\begin{figure*}[t]
\includegraphics[width=0.9\linewidth]{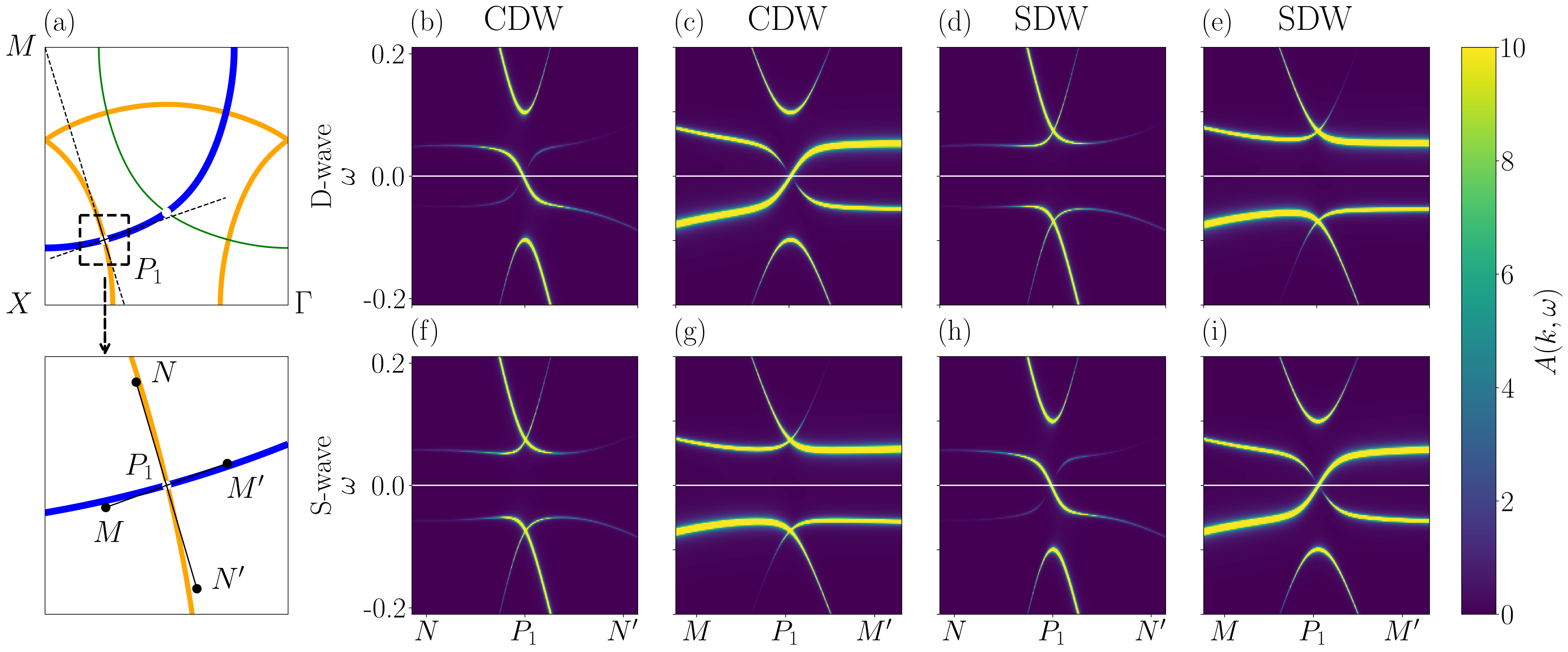}
\caption{Spectral function along two distinct paths across the intersection point $P_1$ for $Q=(\pi / 2,\pi)$. (a) Fermi surfaces in the second quadrant. An enlarged view of the region surrounding $P_1$ is shown in the lower panel. The solid lines represent two paths, one along the radial direction and the other perpendicular to it, crossing $P_1$. The coordinates $(r, \theta)$ of $N$, $N^\prime$, $M$, and $M^\prime$ are $(0.72\pi , 18.1^\circ)$ and $(0.86\pi , 18.1^\circ)$, $(0.79\pi, 15^\circ)$, and $(0.79\pi, 21^\circ)$, respectively.  (b-i) Spectral functions along the two paths for $d$- or $s$-wave pairing states with CDW or SDW couplings. The parameters are $(V,\Delta)=(0.05,0.07)$ for d-wave, and $(V,\Delta)=(0.05,0.05)$ for s-wave, $\eta=0.001$. The white solid line represents the Fermi energy.
 }
\label{fig:Fig2_Q4}
\end{figure*}

 In the above discussion, we have primarily focused on the d-wave pairing state. However, a significant contrast is observed when the superconducting pairing changes from $d$- to isotropic $s$-wave symmetry. In the latter case, there is no sign change between the gap parameters of the main band and the first subband at $P_1$. Consequently, the hybridization of the two bands has a minimal effect on the spectral functions at $P_1$, as shown in Fig.~\ref{fig:Fig2_Q4}f -g. This is in stark contrast to the d-wave pairing case, where the interference effect leads to the complete closure of the energy gap at $P_1$, as depicted in Fig.~\ref{fig:Fig2_Q4}b -c for a CDW-coupled system. Therefore, the gap sign change between the two bands, coupled via a CDW or CDW-like interaction, is a crucial factor for a pronounced phase interference effect.

 On the other hand, when the main band and the first subband are coupled by an SDW or SDW-like interaction, a fascinating role reversal of the $d$- and $s$-wave pairing states is observed, as shown in Fig.~\ref{fig:Fig2_Q4}d -e and h -i. In this case, the $s$-wave pairing state, which lacks a gap sign change between the two bands, surprisingly exhibits a strong interference effect, closing the energy gap at $P_1$. This behavior is reminiscent of the $d$-wave pairing state with CDW interactions. Conversely, the $d$-wave pairing state, with a gap sign change between the two bands, does not show a substantial phase interference effect, behaving similarly to the s-wave pairing state with CDW interactions.

 The above results indicate that we can determine the superconducting gap sign structure by measuring the single-particle spectral function at the intersection points of the main and induced bands, provided we know the interactions, such as CDW or SDW, that induce the coupling. Conversely, if we know the pairing symmetry of the system, we can also identify whether the coupling originates from a CDW- or SDW-like interaction by measuring the single-particle spectral function at these intersection points. This finding highlights the significance of the phase interference effect.
 
 This picture of the phase interference effect is also applicable to higher-order intersection points induced by interactions related to $2Q$ or higher-order couplings. However, the effects at these higher-order intersection points are considerably weaker than those at first-order points. Consequently, the superconducting gaps at higher-order intersection points will not be significantly renormalized by the interference effect, making it difficult to differentiate the relative phase of the energy gaps or the type of interaction.

\section{Comparison with experimental results} 
\label{Sec:experiments}

 \subsection{ARPES results}
 \label{Sec:arpes}

Bismuth-based Bi2212 cuprate superconductors are known for their incommensurate modulations in the crystal structure along the $b^\star$ direction \cite{gao_incommensurate_1988, Withers_1988}. These modulations lead to replica bands induced by the lattice superstructure \cite{PhysRevLett.72.2757,osterwalder1995,PhysRevLett.82.2618}. While most research has focused on the primary band and the effects of these superstructure-induced bands, the interaction between the main and induced bands has largely been overlooked.

 In Bi2212, each unit cell contains two $\ce{CuO_2}$ planes. The hybridization of electronic orbitals between these two layers results in bilayer splitting within the band structures, forming bonding and antibonding bands. Additionally, the crystal structure of Bi2212 exhibits a superstructure modulation with a characteristic wave vector $Q=(0.21,0.21) \pi$ (in units of the inverse lattice constant)~\cite{gao_incommensurate_1988}, which creates a series of subbands from these bonding and antibonding bands. Notably, selective hybridizations are observed in Bi2212~\cite{PhysRevB.101.014513}: the hybridizations between the bonding/antibonding band and the antibonding/bonding subbands are strong, whereas the hybridizations between the bonding (similarly antibonding) band and its descendant subbands are significantly weak. The reasons for this specific hybridization behavior remain unclear. Consequently, the superstructure modulation predominantly induces the coupling between bonding and antibonding. 
 
 Therefore, we model the Bi2212 superconductors by the following Hamiltonian,
\begin{equation}
    H  = H_a +H_b +\sum_{k\sigma}\left(Vc_{k+Q,\sigma}^\dagger d_{k,\sigma} +\text{h.c.}\right) ,\label{eq:H_Q10} 
\end{equation}
where $V$ is the effective potential mimicking the superstructure modulations characterized by wave vector $Q$, $H_a$ and $H_b$ are the Hamiltonians for the antibonding and bonding bands, 
\begin{eqnarray}
    H_a & = & \sum_{k\sigma} \xi^a_k c^\dagger_{k\sigma} c_{k\sigma} + \sum_k \left( \Delta_k c^\dagger_{k\uparrow} c^\dagger_{-k\downarrow} + h.c.\right), \\\
    H_b & = & \sum_{k\sigma} \xi^b_k d^\dagger_{k\sigma} d_{k\sigma} + \sum_k \left( \Delta_k d^\dagger_{k\uparrow} d^\dagger_{-k\downarrow} + h.c.\right), 
\end{eqnarray}
$c_{k\sigma}$ and $d_{k\sigma}$ are the electron operators of the antibonding and bonding bands, respectively. The energy dispersions of the bonding and antibonding bands are 
\begin{equation}
    \xi^{a,b}_{k} = \varepsilon_{k} \pm  t_\perp \left[ a_0+\frac{(\cos k_x-\cos k_y)^2}{4} \right] .
\end{equation}
 The superconducting electrons of Bi2212 are assumed $d_{x^2-y^2}$-wave paired, described by the order parameter
 \begin{equation}
      \Delta_k= \frac{1}{2} \Delta \left( \cos k_x-\cos k_y \right) .  \end{equation}  
As the difference in superconducting energy gaps between the bonding and antibonding bands observed by experiments is small~\cite{Ai_2019}, we ignore this difference and assume the two bands have the same gap amplitude $\Delta$.

\begin{figure*}[t]
\includegraphics[width=0.9\linewidth]{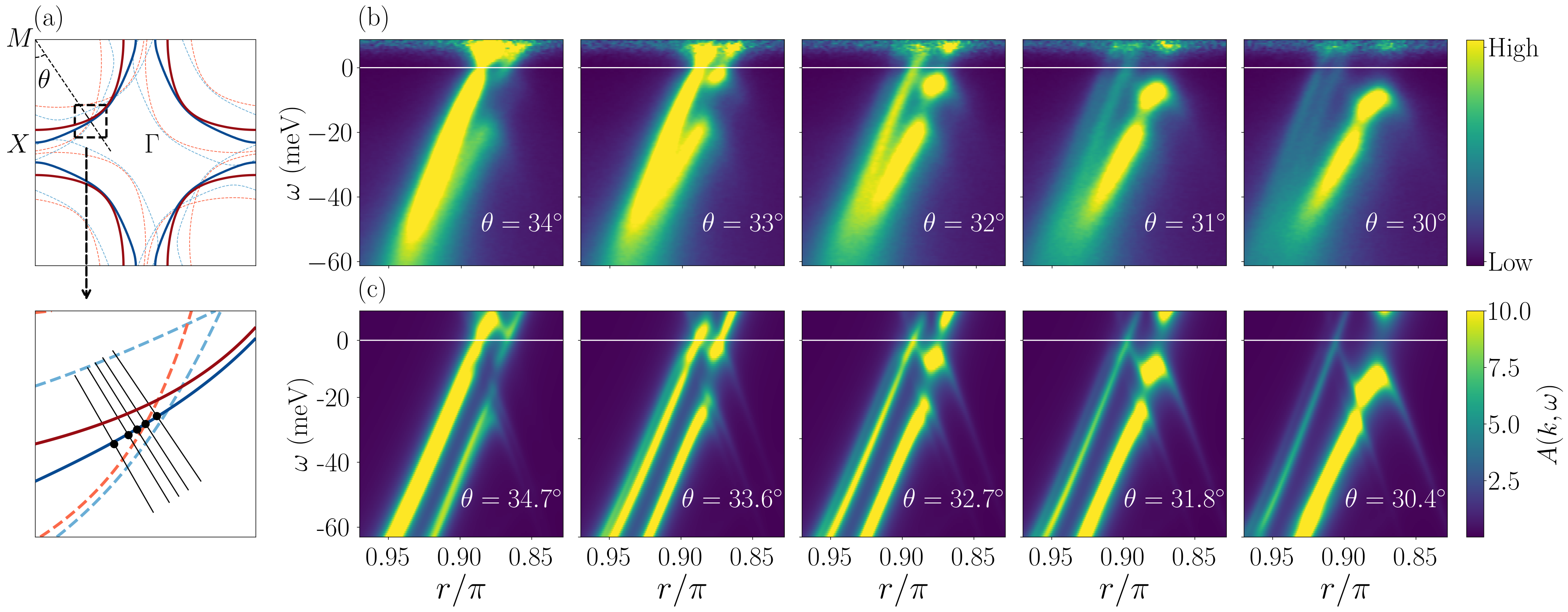}
\caption{
 (a) Normal Fermi surfaces of the bonding (red solid curves) and antibonding (blue solid curves) bands and their first descendent subbands (dashed lines) corresponding to $\pm Q$. The lower panel provides an enlarged view of a small region around an intersection point. (b) Single particle spectral function obtained from ARPES measurements along five radial paths, with $r$ representing the polar distance from the origin at $M=(-\pi, \pi)$, as shown in the lower panel of (a). These paths are indicated by the polar angle $\theta$. (c) Calculated spectral functions along the five paths, whose polar angles are close to the experimental ones shown in (b). The spectral resolution is set to $\eta=0.01$. The white solid line represents the Fermi energy.
}
\label{fig:Fig_Q10}
\end{figure*}

 The model defined by Eq. \eqref{eq:H_Q10} presents a comprehensive framework that surpasses the simple two-band model used in Ref.~\cite{gao_arpes_2024}. It includes a thorough consideration of realistic superstructure modulations, bilayer splitting, and selective hybridizations in Bi2212. To compare with the experimental data, we use $Q=(\pi/5,\pi/5)$ to approximate the incommensurate superstructure wave vector $Q\approx (0.21\pi, 0.21\pi)$ observed in real materials. Additionally, we use the parameters determined by experiments~\cite{gao_arpes_2024} $(V,\Delta)/t_1=(0.033,0.079)$ and $t_1=147.3$ meV. The value of $t_1$ is determined by multiplying the bare hopping integral with the renormalization factor, both extracted from ARPES measurements \cite{gao_arpes_2024}. Other parameters are assigned the values $(t_\perp,a_0,\mu)/t_1=(-0.39,0.095,-1.196)$, as determined by fitting with the band structure of Bi2212~\cite{Markiewicz_2005}. 

Fig.~\ref{fig:Fig_Q10}a shows the Fermi surfaces of the bonding and antibonding bands and their descendent subbands. For clarity, only the Fermi surfaces of the first descendent bands, generated with the $\pm Q$ terms, are depicted. Fig.~\ref{fig:Fig_Q10}b displays the ARPES results for Bi2212 along five radial paths ~\cite{gao_arpes_2024}, denoted by their polar angle $\theta$, shown in the lower panel of Fig.~\ref{fig:Fig_Q10}a. 

Fig.~\ref{fig:Fig_Q10}c shows the calculated spectral function along the five paths with polar angles close to the experimental ones. To be noted, there is a minor discrepancy in the angles of corresponding points (approximately $0.5^\circ$ from the intersection point) when compared to experimental data. This variance arises due to our approximation of the wave vector $Q$,  transitioning from an incommensurate modulation to a commensurate one. Remarkably, the calculated results agree excellently with the ARPES data~\cite{gao_arpes_2024}. Particularly, a gapless behavior is observed near the intersection marked by $\theta=33.6^\circ$, signifying the emergence of new superconducting nodes in addition to the intrinsic $d$-wave gap nodes along two diagonal directions of the Brillouin zone. This serves as clear evidence of the phase interference effect at the intersection point, attributable to the interplay between the $d$-wave energy gap and the superstructure modulation. 

From the calculation, we find that the spectral function appears to lack particle-hole symmetry. This asymmetry stems from a substantial variance in the spectral weight distribution relative to the Fermi energy near the intersection point, exemplified at $\theta=32.7^\circ$. Nevertheless, the energy dispersion remains symmetric with respect to the Fermi energy, as the Hamiltonian does not break the particle-hole symmetry. This extraordinary behavior persists over a relatively large area, approximately 3 degrees around the intersection points, which is promising for experimental detection.

Away from the intersection points, typical $d$-wave gap structures become apparent. A gapless structure emerged at an angle $\theta=30.4^\circ$, seemingly far from the intersection point, which looks abnormal. This anomaly arises because $\theta=30.4^\circ$ aligns closely with the nodal direction of the first subband Fermi surface.

We have also calculated the spectral function near the intersection point, assuming the gap parameter has other pairing symmetries. Specifically, we examined three distinct s-wave pairing states, defined by the gap functions $\Delta_k = \Delta / 2$, $\Delta_k = \Delta (\cos k_x + \cos k_y) / 2$, and $\Delta_k = \Delta \vert \cos k_x - \cos k_y \vert / 2$. However, none of these pairing states reproduce even the qualitative features of the experimental data. This discrepancy provides further evidence in favor of the d-wave pairing symmetry in Bi2212. Detailed discussions are presented in the Appendix.

\subsection{Twisted junctions}
\label{Sec:twisted}

\begin{figure}[t]
\includegraphics[width=0.9\linewidth]{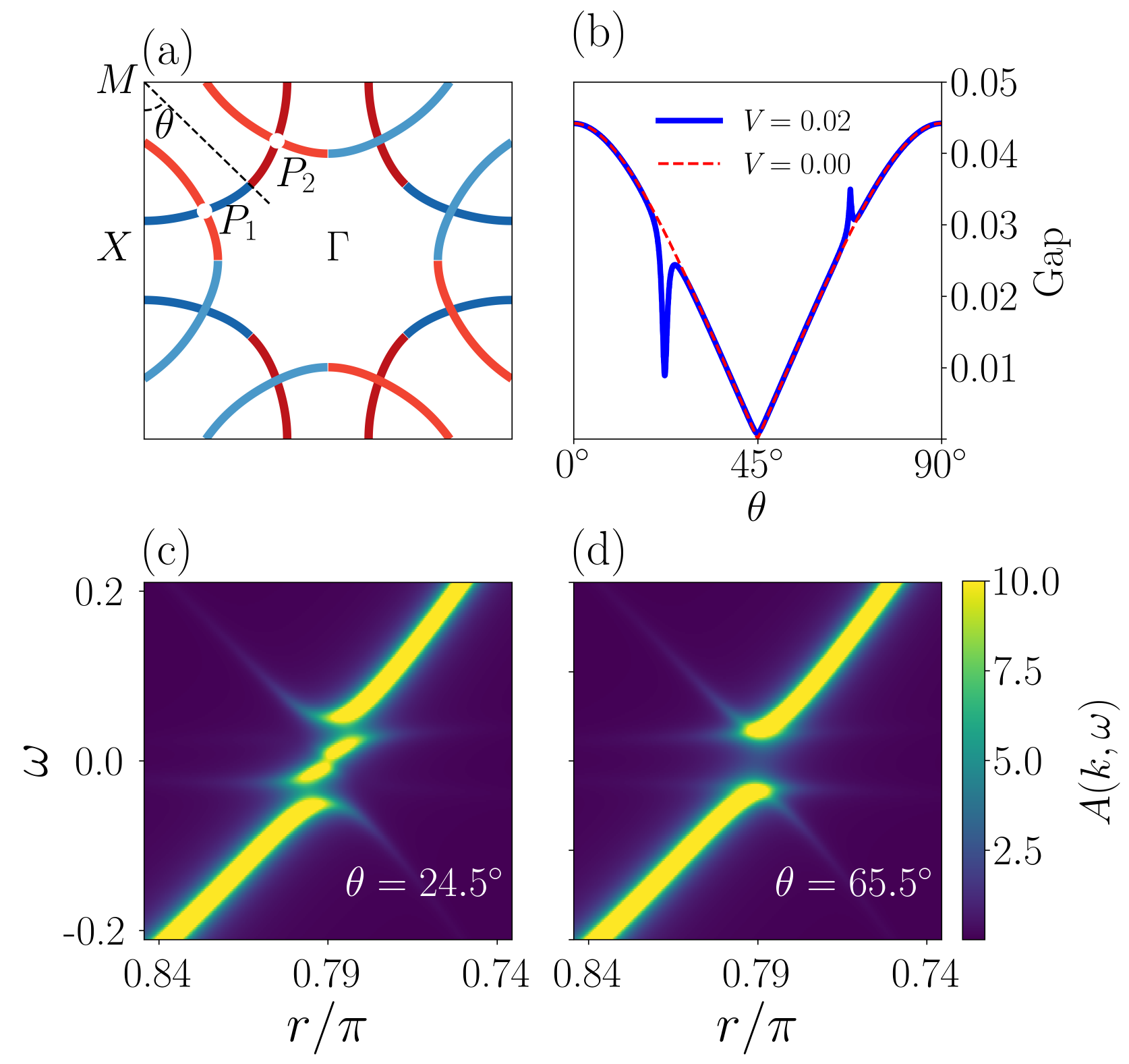}
\caption{
 (a) Fermi surfaces of the twisted junction, with red indicating a positive gap phase and blue indicating a negative gap phase. Points $P_1$ at $(r,\theta)=(0.79\pi,24.5^\circ)$ and $P_2$ at $(0.79\pi,65.5^\circ)$ represent two distinct types of intersection points. At $P_1$, the $d_{x^2-y^2}$-wave energy gaps of the two bands change signs, whereas at $P_2$, they retain the same sign. (b) Absolute values of the gap parameter (blue solid line) as a function of the polar angle along the Fermi surface of the top band in the second quadrant. The red dashed line shows the $d$-wave gap size without interlayer coupling. (c-d) Single-particle spectral functions along two polar paths crossing $P_1$ and $P_2$, respectively. The parameters used are $(V, \Delta, \eta)=(0.02, 0.05, 0.01)$ in units of $t_1$.
}
\label{fig:Fig_Q2_twisted}
\end{figure}

The tunneling effect in a twisted Bi2212 junction continues to be a focal point of interest and debate in high-T$_c$ superconductivity. Recent advancements in van der Waals stacking techniques~\cite{VanderWaals,science.1244358,Gomez_2014,masubuchi_2018} now enable atomically precise creation of interfaces and accurate control of twist angles, which facilitates precise modulation of moire superlattices and interlayer hybridization effects in engineered heterostructures. The Fermi surfaces in twisted junctions exhibit intersection points analogous to those observed in density wave states, thereby manifesting pronounced interference effects under conditions where interlayer hybridization assumes a role equivalent to density wave interactions.

In weakly and incoherently coupled $d$-wave superconductor junctions, theoretical predictions suggest that the tunneling current should vanish, at least to the leading order approximation, if the twist angle is exactly $45^\circ$. However, not all experimental measurements align with this prediction. For instance, an experimental study by Tachiki's group~\cite{PhysRevB.65.140513} reported that the tunneling current is zero within measurement errors in cross-whisker junctions of bismuth-based cuprates at the twist angle of $45^\circ$. In contrast, studies by Xue's group~\cite{PhysRevX.11.031011} and Kim's group~\cite{science.abl} found a finite tunneling current at this twist angle in ultra-thin twisted flakes of bismuth-based cuprates, leading them to suggest that the superconducting pairing in Bi2212 may not have simple $d$-wave symmetry. To explain these discrepancies, several theoretical models have been proposed, including novel tunneling mechanisms~\cite{volkov2021josephson} and the possibility of a time-reversal symmetry-breaking topological superconducting $d+id$ phase~\cite{science.abl}.

Below, we apply the superconducting phase interference effect theory to a twisted junction system of $d$-wave superconductors, focusing on a $45^\circ$ twisted junction. The conclusions drawn here are generally applicable and do not depend solely on the twist angle chosen. Our calculations indicate that the phase interference effect can significantly alter the gap structure of $d$-wave pairing order parameters in a $45^\circ$ twisted junction induced by coherent tunneling. The degree of coherent tunneling is susceptible to the preparation of the samples, especially the quality of the interfaces, which can vary between different experimental groups.

 Let us consider a $45^\circ$-twisted bilayer system, described by the Hamiltonian
 \begin{equation}
    H =  H_0 + \tilde{H}_0 + \sum_{k \tilde{k}  \sigma} \left( V_{k,\tilde{k} } c_{k \sigma}^\dagger d_{\tilde{k} \sigma}+ h.c.\right) ,    \label{eq:H_twisted} 
\end{equation}
 where $c_{k\sigma}$ and $d_{\tilde{k}\sigma}$ are the electron annihilation operators on the top and bottom layers, respectively. $H_0$ is the Hamiltonian of the top layer, described by Eq. \eqref{eq:H0}. $\tilde{H}_0$ is the corresponding Hamiltonian of the bottom layer, similarly defined as $H_0$, but in the framework where the momentum point $\tilde{k}$ is defined by rotating $45^\circ$ with respect to the top layer: 
\begin{equation}
    \tilde{H}_0 =\sum_{\tilde{k}\sigma}  \varepsilon_{\tilde k} d_{\tilde k\sigma}^\dagger d_{\tilde k\sigma} + \sum_{\tilde k} \left(\Delta_{\tilde k} d_{\tilde k\uparrow}^\dagger d_{-\tilde k\downarrow}^\dagger+\text{h.c.}\right) . \label{eq:tildeH0}
\end{equation}

 The rightmost term in Eq. \eqref{eq:H_twisted} is the tunneling term between the two layers with $V_{k \tilde k}$ the effective tunneling matrix element. If the tunneling is completely incoherent, $V_{k \tilde k}=V$, independent of $k$ and $\tilde k$. This corresponds to the case of atomic-scale disorder, the Josephson tunneling current should vanish due to the $d$-wave pairing symmetry of the gap order parameters~\cite{volkov2021josephson}. 
 
 However, if the tunneling is coherent so that the momentum is conserved in the tunneling process
 \begin{equation}
     V_{k \tilde k}= V \delta_{{\tilde k}_x, (k_x-k_y)/\sqrt{2}} \, \delta_{{\tilde k}_y, (k_x+k_y)/\sqrt{2}},
 \end{equation}
 a significant phase interference effect may happen, which will change the pairing gap structures and consequently the tunneling effect. 

 Fig.~\ref{fig:Fig_Q2_twisted}a illustrates the Fermi surfaces of the twisted bilayer system. There are eight intersection points, which can be divided into two groups. The first group includes $P_1$ and three other points generated by sequentially rotating $P_1$ by $90^\circ$. The second group comprises the remaining four points. The $d_{x^2-y^2}$-wave pairing gaps on the two layers change signs at the first group of points but have the same sign at the second group of points. Coherent tunneling introduces a hybridization between the two bands, leading to phase interference effects. These effects differ significantly at $P_1$ and $P_2$ due to the distinct gap sign structures at these points.

 Near the intersection points, as shown in Fig.~\ref{fig:Fig_Q2_twisted}b -d, the gap structure markedly deviates from the standard $d$-wave function. At $P_1$, the gap size is dramatically suppressed. However, at $P_2$, the gap size is slightly enlarged.  This asymmetry removes the reflection symmetry of the gap size on the two sides of the gap node on each segment of the Fermi surface. It provides an alternative framework for probing the relative superconducting phase through twisted superstructures.

 The gap phase interference discussed above relies critically on coherent tunneling. Incoherent coupling across the junction suppresses this phase interference entirely. Enhanced interlayer coupling in pristine interfaces promotes coherent tunneling, while weakened coupling dominated by disorder-induced scattering results in incoherent tunneling. Consequently, successful observation of this effect requires atomically precise junction engineering with minimal interfacial disorder.
 
 In addition to the phase interference effect induced by coherent tunneling, interference between the main band and sub-bands within each Bi2212 flake, as discussed in \ref{Sec:arpes}, can introduce additional gap nodes in the second and fourth quadrants of the Brillouin zone in the twisted junction. These additional nodes further disrupt the reflection symmetry in the gap function.

The above discussion suggests that phase interference can significantly alter the gap structure in twisted Josephson junctions dominated by coherent tunneling, highlighting the delicate interplay between band structure and tunneling dynamics. Whether this effect can fully account for the finite Josephson current observed in $45^\circ$-twisted Bi2212 junctions within a pure $d$-wave pairing scenario remains an open question. Nevertheless, it provides a novel framework for exploring the mechanisms behind this intriguing experimental observation.

\section{Summary}
\label{Sec:summary}

 In summary, we have introduced the superconducting phase interference effect in momentum space. This effect relies on the interplay between two bands that become hybridized through interaction at an energy scale significantly lower than the Fermi energy but comparable to the superconducting energy gap. This hybridization occurs when a density wave, such as CDW, SDW or PDW, a superstructure modulation, or other orders, couples with a primary band, generating a series of replica bands from the primary bands. The coupling of superconducting electrons between the primary band and the induced replica bands creates the interference effect, which can be used to probe the pairing gap phase by measuring the single-particle spectral function. As the periodicity of density waves increases, so does the number of subbands, resulting in numerous intersection points. However, the most significant effects arise from the main band and the first subbands generated by hybridization terms with wave vectors Q and -Q. Higher-order wave vectors have a considerably weaker impact. 

 Hybridization can also occur if two distinctive but intersecting Fermi surfaces are coupled weakly but primarily by momentum-preserved coherent hopping. This may happen, for example, in a tunneling junction formed by two twisted flakes of Bi2212 superconductors. In the case the twisted angle is $45^\circ$, the phase interference effect can modify the $d$-wave pairing gap structure and break the symmetry in the gap amplitude on the two sides of the gap nodes on the Fermi surface.

The phase interference effect is most pronounced at the intersection points of the two hybridized bands. If the hybridization is induced by CDW, PDW-\RNum{2}, or an interaction that relies solely on the charge degree of freedom, such as superlattice modulation, the phase interference effect creates a gapless excitation at the intersection point when the pairing order parameters of the two bands have opposite signs. Conversely, if the hybridization is induced by SDW, PDW-\RNum{1}, or SDW-like interactions, a gapless excitation emerges at the intersection point when the pairing order parameters of the two bands have the same sign. This phase interference effect applies to systems with incommensurate density waves or superstructure modulations. It can help distinguish different types of density waves if the pairing phase at the intersection points is known. Conversely, knowing the types of density waves or modulations, we can determine the gap signs at the intersection points.

Our theory of the phase interference effect provides a good account for the experimental observation of ARPES spectra around intersection points in the Bi2212 superconductor with incommensurate superstructure modulation. It confirms the claim that cupreate superconducting pairing has $d$-wave pairing symmetry. We predict the gap structures near intersection points in systems with $\sqrt{2}a \times \sqrt{2}a$ charge modulation. This prediction can be examined in systems with this type of charge ordering or instability, for example, in overdoped Bi2201. 

This work investigates the superconducting phase interference effect in momentum space, offering a compelling method to study superconducting gap phase structures using momentum-resolving experimental tools like ARPES. It introduces a novel and broadly applicable framework for analyzing not just the phase structures of superconducting order parameters, as discussed in this work, but also other order parameters in other macroscopically ordered phases, again by hybridizing a primary band with an induced replica band. The phase interference effect significantly alters the single-particle spectral function at intersection points, with potential implications for probing other dynamic response functions. While we have applied this theory to cuprates, it is important to emphasize its broader applicability to other materials via band folding mechanisms in superstructures or surface reconstructions, such as in twisted graphene and substrate-controlled film growth systems. 

In underdoped cuprate superconductors, the pseudogap phase emerges in the normal state above $T_c$, leading to a fragmented Fermi surface characterized by disconnected arcs. Nevertheless, in the superconducting phase, ARPES reveals the existence of quasiparticle coherence peaks across the entire Fermi surface, even though the peaks near the antinodal regions are noticeably broader than those near the gap nodes~\cite{Vishik_2010}. This suggests that the interference effects discussed in this work hold, at least qualitatively, even in the presence of the pseudogap at low temperatures, so long as the intersection points do not lie near the antinodal points where the pseudogap is largest. However, elucidating the pseudogap’s quantitative influence on superconducting phase interference effects remains elusive until the origin of the pseudogap is fully understood.

 Our findings suggest that the interference effect offers a useful probe of the relative phase structure near the intersection points of the Fermi surface. However, this effect alone is insufficient to fully resolve the momentum dependence of the superconducting gap phase across the entire Fermi surface. Moreover, our analysis is based on the mean-field approximation, which holds when quantum and thermal fluctuations of the superconducting order parameter are small relative to the order parameter itself. This condition is generally met at lower temperatures and away from quantum critical points. In some materials, density wave and superconducting states can also compete, potentially leading to scenarios where one state significantly suppresses the other. Consequently, self-consistent calculations are crucial for accurately capturing the band hybridization. Further experimental and theoretical studies are needed to deepen our understanding and fully exploit the potential of this effect.

\section*{Competing interests}
The authors declare no competing interests.

\section*{Acknowledgments}

We thank Zixiang Li for helpful discussions. This work is supported by the National Natural Science Foundation of China(12488201) and the National Key Research and Development Program of China (2021YFA1401800).

\section*{Author contributions}
Tao Xiang, Xingjiang Zhou and Bo Zhan proposed and designed the research. Xingjiang Zhou and Qiang Gao provided the experimental data. Bo Zhan conducted the model calculations. Tao Xiang, Bo Zhan and Runze Chi wrote the paper. All authors discussed the results and commented on the manuscript.

\bibliographystyle{SciBull} 
\bibliography{ref}

\begin{appendix}
\label{Appendix}

\section{Spectral functions in Bi2212 assuming s-Wave pairing symmetry}

In this appendix, we present calculated spectral functions near the intersection point in Bi2212, assuming the superconducting gap follows an s-wave rather than a d-wave pairing symmetry. We compare these results with experimental ARPES data. Specifically, three representative s-wave pairing states are studied: (a) an isotropic s-wave state, $\Delta_k = \Delta / 2$; (b) an extended s-wave state, $\Delta_k = \Delta (\cos k_x + \cos k_y) / 2$; and (c) an anisotropic s-wave state, where the gap exhibits nodes along the diagonal directions, similar to a d-wave gap but without any sign change, $\Delta_k = \Delta |\cos k_x - \cos k_y| / 2$.

\begin{figure*}[th]
\centering
\includegraphics[width=0.8\linewidth]{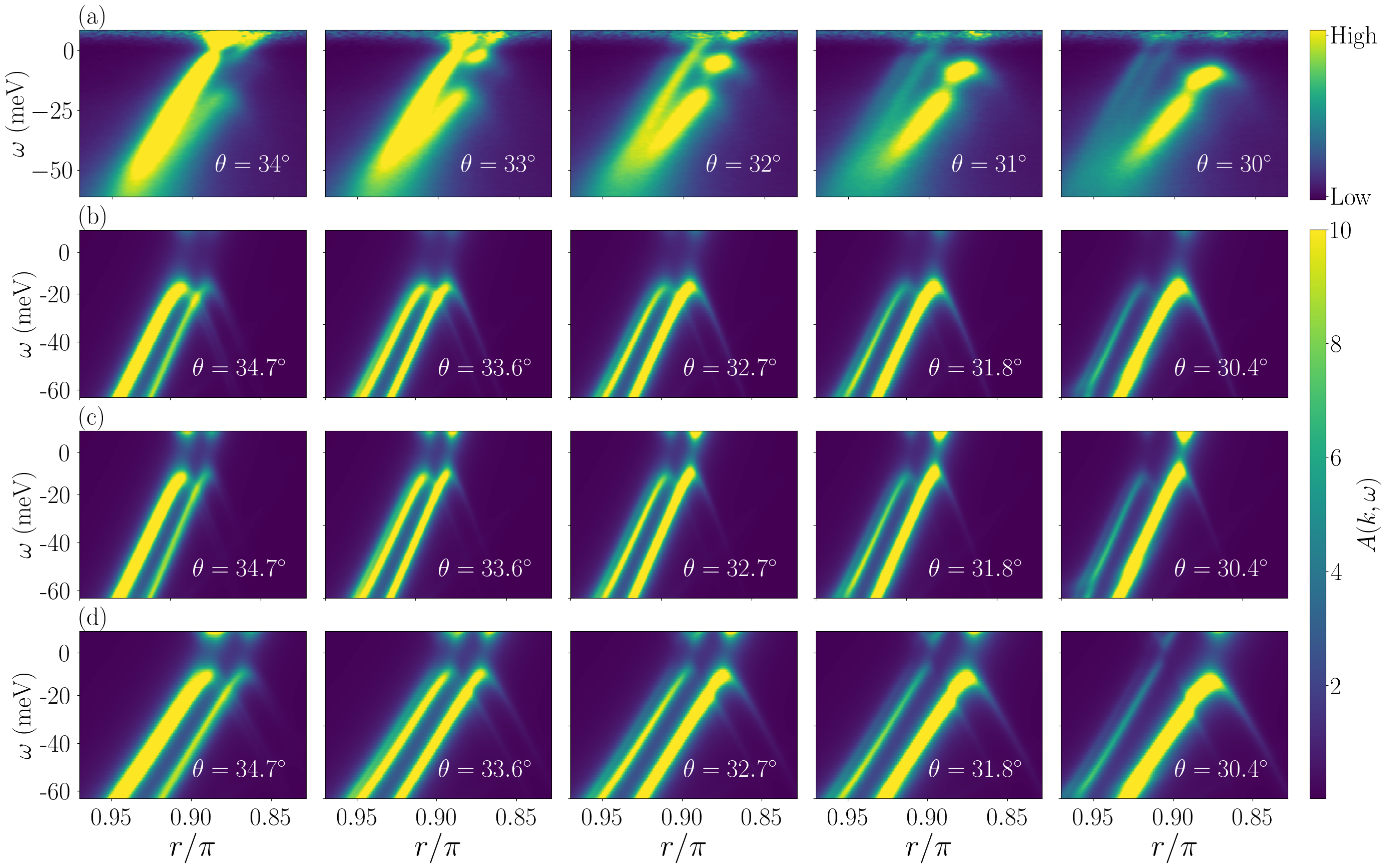}
\caption{ Comparison of the experiment data (a) with the spectral functions calculated for three $s$-wave pairing states: (b) isotropic s-wave,  $\Delta_k = \Delta / 2$, (c) extended s-wave, $\Delta_k = \Delta (\cos k_x + \cos k_y) / 2$, and (d) anisotropic s-wave, $\Delta_k = \Delta |\cos k_x - \cos k_y| / 2$. The same parameters and selected paths as in Fig.~\ref{fig:Fig_Q10} are used. }
\label{fig:Fig_Q10_appendix}
\end{figure*}

As shown in Fig.~\ref{fig:Fig_Q10_appendix}, none of these three s-wave pairing states adequately reproduce the key spectral features observed by ARPES near the intersection point, even at a qualitative level. This comparison suggests that the superconducting electrons in Bi2212 are unlikely to be s-wave paired.

\end{appendix}

\end{document}